\documentclass{article}%
\usepackage{amsmath}
\usepackage{amsfonts}
\usepackage{amssymb}
\usepackage{graphicx}%
\providecommand{\U}[1]{\protect\rule{.1in}{.1in}}

\begin{document}

\begin{center}
\textbf{On the optical-mechanical analogy in general relativity}

Kamal K. Nandi and Anwarul Islam

Department of Mathematics, University of North Bengal, Darjeeling (W.B.)
734430, India

\textbf{Abstract}
\end{center}

We demonstrate that the Evans-Rosenquist formulation of the optical mechanical
analogy, so successful in the application to classical problems, also
describes the motion of massless particles in the Schwarzschild field of
general relativity. It is possible to obtain the well known equations for
light orbit and radar echo delay which account for two exclusive tests of
Einstein's field equations. Some remarks including suggestions for future work
are also added.

\textbf{I. INTRODUCTION}

The historical optical-mechanical analogy$^{1,2}$ has recently been cast into
a familiar form by Evans and Rosenquist.$^{3-5}$ This new formulation, based
on Fermat's principle, provides an interesting approach that can be profitably
utilized in the solution of many classical problems. The approach works either
way: The well-known ideas and techniques of classical mechanics can be
successfully applied to the problem of classical optics and vice versa. Such
success naturally prompts further enquiry as to whether the applicability of
the optical-mechanical analogy could be extended even beyond the classical
regimes. More specifically, a curious student might ask: Can the
Evans-Rosenquist (ER) formulation be used to describe the phenomenon of light
propagation and massive particle motion in general relativity (GR)? In order
to decide this question, let us note that it addresses two distinct areas of
investigation:One has to first examine if the \textquotedblleft$F=ma$%
\textquotedblright\ optics of ER does at all describe light propagation in the
GR scenario. If it does, then the first hurdle is overcome and the question of
applying the optical analogy to massive particle motion becomes meaningful. To
what extent this analogy would describe planetary motion in GR is a matter of
separate investigation.

In this paper, we shall examine only the first part of the question that
relates to the motion of massless particles in GR. We choose Schwarzschild
field of gravity primarily because of its experimental importance. Besides, on
many occasions, it is used as an ideal example for illustrating fundamental
notions of GR.

The exterior Schwarzschild metric, which is a solution of Einstein's GR field
equations, has been astoundingly successful in describing various
gravitational phenomena. Famous experimental tests of GR include the bending
of light rays, perihelion advance of a planet, radar echo delay, and
gravitational red shift in the environment of the Schwarzschild gravity field
generated by a static, spherically symmetric material source like the
sun.$^{6}$

In the present approach, the only ingredient of GR to be used is the isotropic
form of the metric relevant to the gravity field (here the Schwarzschild
field). No sophisticated mathematics involving four vectors, tensors,
Christoffel symbols, or geodesic equations will be necessary. In addition to
being simpler and hence understandable by a wide range of readers, the ensuing
exposition offers a different window through which the GR events can be
visualized. From the instructional point of view, it is always useful to look
at familiar things from as many different perspectives as possible.

It would be worthwhile to display the relevant formulas of the
optical-mechanical analogy that we are going to use in our paper. This is done
in Sec.II. Thereafter, as a necessary step, the Schwarzschild gravity field is
portrayed as a refractive optical medium in Sec.III. GR equations for the
light orbit and the radar echo delay in Schwarzschild gravity are derived in
Secs.IV and V respectively. Finally, some relevant remarks including a few
suggestions for future work are made in Sec.VI.

\textbf{II. OPTICAL-MECHANICAL\ ANALOGY}

We need not go into the detailed formulation of the \textquotedblleft$F=ma"$
optics of ER. Instead, for our present purpose, we reproduce only a table
which is self-explanatory$^{3}$

\bigskip

Quantity
\ \ \ \ \ \ \ \ \ \ \ \ \ \ \ \ \ \ \ \ \ \ \ \ \ \ \ \ \ \ \ \ \ \ \ \ Mechanics
\ \ \ \ \ \ \ \ \ \ \ \ \ \ \ \ \ \ \ \ \ \ \ \ \ \ \ \ \ \ \ \ \ Optics

------------------------------------------------------------------------------------------------

Position
\ \ \ \ \ \ \ \ \ \ \ \ \ \ \ \ \ \ \ \ \ \ \ \ \ \ \ \ \ \ \ \ \ \ \ \ \textbf{\ }%
$\overrightarrow{r}(t)$%
\ \ \ \ \ \ \ \ \ \ \ \ \ \ \ \ \ \ \ \ \ \ \ \ \ \ \ \ \ \ \ \ \ \ \ \ \ \ \ \ \ \ \ \ $\overrightarrow
{r}(a)$

\textquotedblleft Time"
\ \ \ \ \ \ \ \ \ \ \ \ \ \ \ \ \ \ \ \ \ \ \ \ \ \ \ \ \ \ \ \ \ \ \ \ \ \ \ \ $t$%
\ \ \ \ \ \ \ \ \ \ \ \ \ \ \ \ \ \ \ \ \ \ \ \ \ \ \ \ \ \ \ \ \ \ \ \ \ \ \ \ \ \ \ \ \ \ \ \ \ \ $a$%

\textquotedblleft velocity"
\ \ \ \ \ \ \ \ \ \ \ \ \ \ \ \ \ \ \ \ \ \ \ \ \ \ \ \ \ \ \ \ \ \ \ $\frac
{d\overrightarrow{r}}{dt}\equiv\overset{.}{\overrightarrow{r}}$
\ \ \ \ \ \ \ \ \ \ \ \ \ \ \ \ \ \ \ \ \ \ \ \ \ \ \ \ \ \ \ \ \ \ \ \ \ \ \ $\frac
{d\overrightarrow{r}}{da}\equiv\overrightarrow{r}^{\prime}$

\textquotedblleft Potential energy"
\ \ \ \ \ \ \ \ \ \ \ \ \ \ \ \ \ \ \ \ $U(\overrightarrow{r})$
\ \ \ \ \ \ \ \ \ \ \ \ \ \ \ \ \ \ \ \ \ \ \ \ \ \ \ \ \ \ \ \ \ \ \ \ \ \ \ \ \ $-n^{2}%
(\overrightarrow{r})/2$

\textquotedblleft Mass"
\ \ \ \ \ \ \ \ \ \ \ \ \ \ \ \ \ \ \ \ \ \ \ \ \ \ \ \ \ \ \ \ \ \ \ \ \ \ \ $m$
\ \ \ \ \ \ \ \ \ \ \ \ \ \ \ \ \ \ \ \ \ \ \ \ \ \ \ \ \ \ \ \ \ \ \ \ \ \ \ \ \ \ \ \ \ \ \ \ $1$%

\textquotedblleft Kinetic energy"
\ \ \ \ \ \ \ \ \ \ \ \ \ \ \ \ \ \ \ \ \ \ \ $T=\frac{m}{2}\left\vert
\overset{.}{\overrightarrow{r}}\right\vert ^{2}$
\ \ \ \ \ \ \ \ \ \ \ \ \ \ \ \ \ \ \ \ \ \ \ \ \ \ \ \ \ \ \ \ $\frac{1}%
{2}\left\vert \overrightarrow{r}^{\prime}\right\vert ^{2}$

\textquotedblleft Total energy"
\ \ \ \ \ \ \ \ \ \ \ \ \ \ \ \ \ \ \ \ \ \ \ \ \ \ $\frac{m}{2}\left\vert
\overset{.}{\overrightarrow{r}}\right\vert ^{2}+U$
\ \ \ \ \ \ \ \ \ \ \ \ \ \ \ \ \ \ \ \ \ \ \ \ \ \ \ \ \ \ \ \ \ $\frac{1}%
{2}\left\vert \overrightarrow{r}^{\prime}\right\vert ^{2}-n^{2}/2$

\textquotedblleft Equation of motion" \ \ \ \ \ \ \ \ \ \ \ \ \ \ \ $m\overset
{..}{\overrightarrow{r}}=-$grad$U$
\ \ \ \ \ \ \ \ \ \ \ \ \ \ \ \ \ \ \ \ \ \ \ \ \ \ \ $\overrightarrow
{r}^{\prime\prime}=$grad$\left(  \frac{n^{2}}{2}\right)  $

------------------------------------------------------------------------------------------------

As can be seen, the role of time $t$ is played by the ER stepping parameter
$a$ having the dimension of length so that $\overrightarrow{r}^{\prime}$ is
not a velocity. It is a dimensionless quantity. The transition between $t$ and
$a$ can be accomplished by using the relation%
\begin{equation}
da=\frac{c_{0}}{n^{2}}dt,
\end{equation}
where $n$ denotes the refractive index of the optical medium, $c_{0}$ denotes
the vacuum speed of light. Evans$^{4}$ has shown that the stepping parameter
can be physically interpreted as \textquotedblleft optical action". Just as a
mechanical particle progresses in time along its trajectory, light progresses
in optical action along its ray.

The identification%
\begin{equation}
\ U(\overrightarrow{r})\ =-\frac{n^{2}}{2}%
\end{equation}
giving the \textquotedblleft potential" $U$ essentially bestows a mechanical
character to photon motions; the only restriction being that the motion
corresponds to mechanics at \textquotedblleft zero total energy." We can also
imagine the possibility that, at least in the classical regime, optical and
mechanical motions can take place under the same form of "force"/force law. An
excellent example is probably the Luneburg lens in optics and the harmonic
oscillator in mechanics$^{3}$. In order to use the ER formulation in the GR
regime, it would be necessary to associate a single scalar function, the
optical refractive index $n$, with the gravity field under consideration. The
"force" on the massless particles moving in that gravity field can then be
explicitly obtained$^{7}$ via the ER expression grad$\left(  \frac{-n^{2}}%
{2}\right)  $.

From now on, we shall use primes ($^{\prime}$) only to designate a radial
coordinate ($r^{\prime}$) and not differentiation with respect to the stepping
parameter $a$. All the differentiations will be displayed in full.

\begin{center}
\textbf{III. GRAVITY FIELD AS A REFRACTIVE MEDIUM}
\end{center}

One might wonder what relationaship could there possibly be between two
entities apparently as diverse as a gravity field and a refractive optical
medium? But, indeed, there is one! It was guessed by Einstein and
Eddington$^{8}$, formally developed by Plebanski and others$^{9}$ and utilized
in the investigation of of specific problems by many physicists.$^{9,10}$ We
shall only quote the results in a form that is easily intelligible.
Plebanski$^{9}$ has shown that, in a curved spacetime with a metric tensor
$g_{\alpha\beta}$, Maxwell's electromagnetic equations can be rewritten as if
they were valid in a flat spacetime in which there is an optical medium with a
constitutive equation. More specifically, with regard to light propagation,
the gravity field behaves as a refractive optical medium. For example, the
gravitational field exterior to a static, spherically symmetric mass $M$ is
equivalent to an isotropic, non-homogeneous optical medium with a refractive
index $n$ given by$^{10}$%
\begin{equation}
n^{2}(r)=\left(  1+\frac{m}{2r}\right)  ^{6}\left(  1-\frac{m}{2r}\right)
^{-3},\text{ \ \ \ \ \ \ }m=GM/c_{0}^{2}%
\end{equation}
where, as usual, $G$ is the gravitational constant, $r$ is the Euclidean
radial coordinate.

Advanced students who are likely to have a fair grasp on the two field
theories, Einstein's and Maxwell's, along with the details of algebraic
manipulations should have no difficulty in pursuing the analysis of Plebanski
and de Felice$^{9,10}$. However, there is a simpler alternative derivation of
Eq.(3), given below, that demands only the knowledge of the form of the
Schwarzschild exterior metric. One may also take Eq.(3) at its face value and
proceed by treating this $n(r)$ as just a given choice of the refractive index
in the ER\ formulation. We shall essentially investigate the consequences of
such a choice in the succeeding sections.

Consider the exterior Schwarzschild metric in standard coordinates
($r^{\prime},\theta,\varphi,t$)%
\[
ds^{2}=\left(  1-\frac{2m}{r^{\prime}}\right)  c_{0}^{2}dt^{2}-\left(
1-\frac{2m}{r^{\prime}}\right)  ^{-1}dr^{\prime2}-r^{\prime2}(d\theta^{2}%
+\sin^{2}\theta d\varphi^{2})
\]%
\begin{equation}
=B(r^{\prime})c_{0}^{2}dt^{2}-A(r^{\prime})dr^{\prime2}-r^{\prime2}%
(d\theta^{2}+\sin^{2}\theta d\varphi^{2})
\end{equation}
where $r^{\prime}>2m$. Redefine the radial coordinate as $r^{\prime}%
=r\phi^{-1}(r)=r\left(  1+\frac{m}{2r}\right)  ^{2}$, so that, if $r=u^{-1}%
$and $r^{\prime}=u^{\prime-1}$, then
\begin{equation}
u^{\prime}=u\phi(u),
\end{equation}%
\begin{equation}
\phi(u)=\left(  1+\frac{mu}{2}\right)  ^{-2}.
\end{equation}

The metric (4) can then be rewritten in the so called isotropic coordinates
($r,\theta,\varphi,t$) as%
\begin{equation}
ds^{2}=\Omega^{2}(r)c_{0}^{2}dt^{2}-\phi^{-2}(r)[dr^{2}+r^{2}(d\theta^{2}%
+\sin^{2}\theta d\varphi^{2})]=\Omega^{2}(r)c_{0}^{2}dt^{2}-\phi
^{-2}(r)\left\vert d\overrightarrow{r}\right\vert ^{2},
\end{equation}
where $\Omega(r)=\left(  1+\frac{m}{2r}\right)  ^{-1}\left(  1-\frac{m}%
{2r}\right)  .$

The above form of the metric has a conformally Euclidean spatial part.$^{11}$
Therefore, the isotropic coordinate speed of light $c(r)$ at any arbitrary
point in the gravitational field, obtained by putting $ds^{2}=0$, is%
\begin{equation}
c(r)=\left\vert \ \frac{d\overrightarrow{r}}{dt}\right\vert =c_{0}%
\phi(r)\Omega(r).
\end{equation}
Defining the refractive index as $n(r)=c_{0}/c(r)$, we have%
\begin{equation}
n(r)=\Omega^{-1}\phi^{-1}=\left(  1+\frac{m}{2r}\right)  ^{3}\left(
1-\frac{m}{2r}\right)  ^{-1},
\end{equation}
which is precisely the same as Eq.(3) above. It is easy to see that%
\begin{equation}
\phi^{2}(u^{\prime})=2^{-4}\left[  1+(1-2mu^{\prime})^{1/2}\right]  ^{4},
\end{equation}%
\begin{equation}
\Omega^{2}(u^{\prime})=1-2mu^{\prime}%
\end{equation}
and%
\begin{equation}
du^{\prime}=\Omega(u)\phi(u)du.
\end{equation}
All the above relationships will be used throughout the paper. It should be
mentioned that Sj\"{o}din$^{12}$ has also derived the expression for $\phi$,
$\Omega$ and $n$ by a different method based on Rindler's operational
definitions of length and time in a gravitational field.

The refractive index $n(r)$ above corresponds to a radial "force law" having a
magnitude%
\[
\text{\textquotedblleft}F"=\frac{dU}{dr}=\frac{m(m-4r)(m+2r)^{5}}%
{8(m-2r)^{3}r^{5}}=\frac{2m}{r^{2}}+\frac{15m^{2}}{2r^{3}}+\frac{27m^{3}%
}{2r^{4}}+O(m^{4}).
\]

\begin{center}
\textbf{IV. LIGHT ORBIT EQUATION}
\end{center}

Within the ER framework, the optical analog of the mechanical zero total
energy or in short \textquotedblleft zero total energy" (see the entry in the
table in Sec.II) is given by%
\begin{equation}
\frac{1}{2}\left\vert \ \frac{d\overrightarrow{r}}{da}\right\vert ^{2}%
-\frac{n^{2}}{2}=0.
\end{equation}
We straightaway claim that this very equation is the GR\ light orbit equation
in Schwarzschild gravity, provided $n$ is given by Eq.(9). In order to see
that, we proceed as follows. Since there is spherical symmetry in the problem,
we can, without loss of generality, fix a plane in which the orbiting has to
take place. It is customary$^{13}$ to choose $\theta=\pi/2$. Writing out
Eq.(13) in plane polar coordinates, we have%
\begin{equation}
\left(  \frac{dr}{da}\right)  ^{2}+r^{2}\left(  \frac{d\varphi}{da}\right)
^{2}-n^{2}(r)=0.
\end{equation}
Noting that Eq.(13) is the first integral of the \textquotedblleft equation of
motion" \ $\overrightarrow{r}^{\prime\prime}=$grad$\left(  \frac{n^{2}}%
{2}\right)  $. Since $n$ does not depend on $\varphi$, the $\varphi$ component
of the equation of motion yields a conserved quantity called by ER as the
\textquotedblleft angular momentum" $h$ given by%
\begin{equation}
h=r^{2}\frac{d\varphi}{da}=\text{ const.}%
\end{equation}
Eliminating the stepping parameter $a$, and writing $r=u^{-1}$, we have%
\begin{equation}
u^{2}+\left(  \frac{du}{d\varphi}\right)  ^{2}-n^{2}h^{-2}=0.
\end{equation}
Evans and Rosenquist have solved a number of problems in classical
optics/mechanics using different choices for $n(r)$. Below we shall discuss,
in the context of GR, some aspects of the Eqs.(13)-(16):

(i) As ER have already discussed, the optical quantity $h$ is completely
different from the corresponding conserved mechanical quantity $h_{0}%
=r^{2}d\varphi/dt$ which is proportional to the areal velocity. In the optical
formalism, we have $h=c_{0}^{2}n^{2}r^{2}d\varphi/dt$ so that $h_{0}\varpropto
n^{-2}$ and $d\varphi/dt\varpropto n^{-2}r^{-2}$. Hence, with our form of
$n(r)$, Eq.(9), neither the areal velocity $h_{0}$ nor the angular velocity
$d\varphi/dt$ remains constant. Of course, for the same force law, the optical
and mechanical zero energy orbits must have the same form, since $a$ or $t$
are ultimately eliminated. We see that the GR light orbit equation has the
form of well known classical optics equation$^{14}$, although this interesting
fact is not widely noticed. The reason for this is that in most text books on
GR, the light orbit equation is given in a completely different form.

(ii) Let us obtain the familiar text book form. Using Eqs. (5), (11) and (12)
in Eq.(16), we get%
\begin{equation}
\phi^{-2}(1-2mu^{\prime})^{-1}\left(  \frac{du^{\prime}}{d\varphi}\right)
^{2}+u^{\prime2}\phi^{-2}-n^{2}h^{-2}=0.
\end{equation}
Using Eq.(9) for $n$, we find%
\begin{equation}
u^{\prime2}+\left(  \frac{du^{\prime}}{d\varphi}\right)  ^{2}-2mu^{\prime
3}-h^{-2}=0.
\end{equation}
Differentiating with respect to $\varphi$, we get%
\begin{equation}
u^{\prime}+\frac{d^{2}u^{\prime}}{d\varphi^{2}}-3mu^{\prime2}=0.
\end{equation}
This is precisely \ light orbit equation in Schwarzschild gravity in standard
coordinates. Therefore, one can say that the familiar Eq.(19) represents in
disguise just the optical analog of the classical zero total energy mechanics.
This conclusion will find a further justification in the development of Sec.V.

(iii) It would be of interest to see how the \textquotedblleft angular
momentum" $h$ is related to the conserved GR quantities $E$ and $L$,
associated with the Killing fields $\partial/\partial t$ and $\partial
/\partial\varphi$, respectively. This would also provide a relationship
between the ER\ stepping parameter $a$ and the geodesic affine parameters used
by relativists. Integration of GR null geodesic equation gives%
\begin{equation}
\left\vert \ \frac{d\overrightarrow{r}}{dt}\right\vert ^{2}=c_{0}^{2}%
\Omega^{2}\phi^{2},\text{ \ \ }\Omega^{2}\frac{dt}{dp}=E,\text{ \ \ }\phi
^{-2}r^{-2}\frac{d\varphi}{dp}=L,
\end{equation}
where $ds^{2}=\lambda dp^{2}$, $\lambda$ is a constant and $p$ is a new
geodesic affine parameter such that $ds^{2}=0\Rightarrow\lambda=0$,
$dp^{2}\neq0$. Eliminating $t$ from Eqs.(20), we get%
\begin{equation}
c_{0}^{-2}E^{-2}\phi^{-4}\left\vert \ \frac{d\overrightarrow{r}}%
{dp}\right\vert ^{2}-n^{2}=0.
\end{equation}
If we now define%
\begin{equation}
dp=c_{0}^{-1}E^{-1}\phi^{-2}da
\end{equation}
then Eq.(13) follows immediately from Eq.(21). We also find that $h=L/c_{0}E$.
From Eqs.(20 and (22), there also follows the ER relation (1) connecting $dt$
and $da$. Further Eq.(22) implies that $ds^{2}=\lambda c_{0}^{-2}E^{-2}%
\phi^{-4}da^{2}$, giving the connection between $a$ and the affine parameter
$s$.

(iv) By integrating Eq.(19), one obtains all allowable light orbits$^{15}$.;
bound, unbound, or even the so-called \textit{boomerang} orbits$^{16}$. There
have been many observations confirming the GR predictions of the bending of
light rays just grazing the sun, the amount being $\Delta\varphi\sim4m/R_{0}$,
where $R_{0}$ is the solar radius. Interested readers may consult any textbook
on GR. We shall only make a relevant remark here. M\O ller$^{17}$ has shown
that the bending of light rays is due partly to the geometrical curvature of
space and partly to the variation of light speed in a Newtonian potential. In
fact, the ratio of the parts is $50:50$. The GR\ null trajectory equations can
be integrated, once assuming a Euclidean space with a variable light speed and
again a curved space with a constant light speed; both contributing just half
the amount of the observed bending. In the present approach, on the other
hand, we are describing light motion by means of a scalar function $n(r)$.
Thus with regard to the light propagation, it looks as if spatial curvature
and Newtonian potential lost their separate identities and merged into an
equivalent refractive medium.There is, of course, no point in asking which one
has a physical reality and which one has not; all these are mathematical
constructs [like the $n(r)$ here] designed only to interpret our physical observations.

(v) Finally, some words of caution. From the similarity of Eqs.(16) and (18),
one might be tempted to conclude that in the $u^{\prime}$ coordinates, $n$ is
given by $n(u^{\prime})=(1+2mh^{2}u^{\prime3})^{1/2}$, but that would be
incorrect. The correct expression for $n(u^{\prime})$ is obtained by using the
fact that $n(u)$ transforms as a scalar. Hence, one obtains$^{18}$%
\begin{equation}
n(u^{\prime})=4(1-2mu^{\prime})^{-1/2}[1+(1-2mu^{\prime})^{1/2}]^{-2}.
\end{equation}
Also, it must be understood that, with this $n(u^{\prime})$, it is not
possible to define an isotropic coordinate velocity of light in the
$u^{\prime}$ coordinates. From a direct calculation with the metric (4), it
will turn out that the coordinate velocities of light are different in radial
and cross radial directions.

\begin{center}
\textbf{V. RADAR ECHO DELAY}
\end{center}

Let us now go beyond the geometrical shape of the light orbit and consider the
dynamics along its path. In other words, we shall derive the GR equations of
motion involving the time $t$ for the propagation of light rays around a
static, spherically symmetric gravitating mass $M$.

Once again we start from the ER \textquotedblleft zero energy" equation
$\frac{1}{2}\left\vert \ \frac{d\overrightarrow{r}}{da}\right\vert ^{2}%
-\frac{n^{2}}{2}=0$ and claim that it is the GR equation we are looking for
provided $n$ is given by Eq.(9). To see this, consider Eqs.(13)-(15) and write%
\begin{equation}
\left(  \frac{dr}{da}\right)  ^{2}+h^{2}r^{-2}-n^{2}(r)=0.
\end{equation}
Utilizing the redefinition $u\rightarrow u^{\prime}$ and the expression for
$n$ from Eq.(9), we have%
\begin{equation}
\phi^{4}u^{\prime-4}\left(  \frac{du^{\prime}}{da}\right)  ^{2}+h^{2}%
(1-2mu^{\prime})u^{\prime2}-1=0.
\end{equation}
At the position $r_{0}^{\prime}[=u_{0}^{\prime-1}]$ of the closest approach to
the gravitating mass $M$, we have $dr^{\prime}/da=0\Rightarrow du^{\prime
}/da\mid_{u_{0}^{\prime}}=0$, giving $h^{2}=(1-2mu_{0}^{\prime})^{-1}%
u_{0}^{\prime-2}$. Putting it in Eq.(25), we find%
\begin{equation}
\phi^{4}\Omega^{-2}u^{\prime-4}\left(  \frac{du^{\prime}}{da}\right)
^{2}+u_{0}^{\prime-2}u^{\prime2}(1-2mu^{\prime})^{-1}-\Omega^{-2}=0.
\end{equation}
Noting from the metric (4) that $B(r^{\prime})=\Omega^{2}$ and $A(r^{\prime
})=\Omega^{-2}$, and using Eq.(1), we finally have%
\begin{equation}
c_{0}^{-2}A(r^{\prime})B^{-2}(r^{\prime})\left(  \frac{dr^{\prime}}%
{dt}\right)  ^{2}+r_{0}^{\prime2}r^{\prime-2}B^{-1}(r_{0}^{\prime}%
)-B^{-1}(r^{\prime})=0.
\end{equation}
This is precisely the textbook form of what is known as the radar echo delay
equation in Schwarzschild gravity.- and, once again, we see that it is still
the same ER \textquotedblleft zero energy" mechanics, only buried in the
($r^{\prime},t$) language. Equation (27) is integrated to obtain the time $t$
required by the light signal (in practice, a radar signal) to travel from one
point of space to another. It is also evident that the coordinate speed of
light $\frac{dr^{\prime}}{dt}$ is less than what it would be if the
gravitating mass were absent. In other words, light is slowed down and the
travel time is longer. All observers will nonetheless measure a photon's speed
to be $c_{0}$ through their positions. In a round trip journey around the mass
$M$, there would be a net GR delay in the radar echo reception. This GR
prediction has been confirmed to a great accuracy by sending radar signals
from Earth to Mercury at superior conjunction and back.$^{19}$

\begin{center}
\textbf{VI. SOME REMARKS}
\end{center}

The contents of the entire paper vividly demonstrate that light propagation in
Schwarzschild gravity is indeed describable by the language of
\textquotedblleft$F=ma"$ optics, a shorthand for the optical mechanical
analogy. It is remarkable that the ER formulation, emerging basically from the
investigations of classical problems, works equally well also in the GR regime.

We are concerned with the predictions for light propagation that follow
exclusively from Einstein's field equations. On the other hand, the formula
for gravitational redshift can be derived from special relativity and the
principle of equivalence alone. The field equations or their solutions need
not be used. In that sense, the redshift is not a prediction following
exclusively from field equations, albeit the effect \textit{does} follow also
from the latter. In the present approach, the usual principle of equivalence
that equates the effects of gravity and artificial forces, is translated into
an equivalence of the effects of gravitation and refractive medium. Using this
idea, the gravitational redshift formula can be obtained in its familiar form.
We shall present the deductions in a separate paper.

It is evident that the considered approach can deal with a reasonably general
class of gravity fields where the metric can ce cast into an isotropic form.
The latter makes it possible to construct a scalar function $n$ which acts as
a representative of the gravity field in he matter of light propagation.
However, the whole range of GR solutions corresponding to different field
distributions can not be made to correspond to such single scalar functions.
At best, a detailed constitutive tensor for the equivalent medium may be
developed$^{10}$. In this case, the challenging task would be to generalize
the ER\ equations appropriately. Nevertheless, we can list some immediate
future works:

(i) The motion of a light pulse inside and across the body of a spherical star
can be tackled quite comfortably. The interior of a spherical star (like the
sun) is described by the Schwarzschild interior metric. It corresponds to a
refractive medium with index, say, $n_{1}$ while the exterior field
corresponds to $n$, Eq.(9). Therefore the whole problem is reduced to one of
light propagation in a composite media with indices $n$ and $n_{1}$. At the
interface, the matching condition is provided by none other than good old
Snell's law. The result will provide a theoretical idea about the total path
of light rays between the sun's core and the Earth.

(ii) For the sake of completeness, we would expect the \textquotedblleft%
$F=ma"$ optics to describe also the motion of massive particles (planetary
motion) in the Schwarzschild gravity field. To that end, efforts are underway
to extend the present treatment. The answer \ to the second part of the
question raised in sec.I would then be available.

Finally, it must be emphasized that there is neither any substitute for nor
shortcut to to the beauty, generality, and richness of Einstein'e general
relativity theory. One must eventually grasp all the details of the physics,
mathematics, and the philosophy of this magnificent never ending edifice. On
the other hand, the language of the ER\ optical-mechanical analogy has the
power to stimulate the interests of a wide cross section of readers who do not
have a formal training in the sophistications of GR. Those who have the
training may, however, regard the preceding developments as providing yet
another avenue to the same experimentally verified tests of light motion in
GR. The educational importance of such alternative points of view cannot be mistaken.

\textbf{ACKNOWLEDGMENTS}

We are deeply indebted to Professor J. Evans for his criticisms and
suggestions that have led to a considerable improvement of the paper. One of
us (A.I.) is grateful to ICCR of the Govt. of India for a Ph.D. Scholarship.

\textbf{REFERENCES}

$^{1}$As quoted by Evans, the optical-mechanical analogy was formulated by
Hamilton (1833) in terms of characteristic functions. See, J. Evans, "The ray
form of Newton's law of motion", Am. J. Phys. \textbf{61}, 347-350 (1993). In
this paper Evans derives the ray form without the use of characteristic functions.

$^{2}$Historically, the basic idea of the optical mechanical analogy was
proposed much earlier by none other than Descartes (1637). This information is
due to J. Arnaud, "Analogy between optical rays and nonrelativistic particle
trajectories:\ A comment", Am. J. Phys. \textbf{44}, 1067-1069 (1976). In this
paper are also discussed the limitations and significance of the analogy.

$^{3}$Essential ingredients of what is referred to as the ER approach are
contained in: J. Evans and M. Rosenquist, ""F=ma" optics", Am. J. Phys.
\textbf{54}, 876-883 (1986).

$^{4}$J. Evans, "Simple forms of equations of rays in gradient-index lenses",
Am. J. Phys.\textbf{ 58}, 773-778 (1990). see especially p.774.

$^{5}$Newton's laws of motion are obtained directly from Fermat's principle,
in M. Rosenquist and J. Evans, "The classical limit of quantum mechanics from
Fermat's principle and the de Broglie relation", Am. J. Phys. \textbf{56},
881-882 (1988).

$^{6}$For a general information: Latest confirmations of GR come from the
observations of the Hulse-Taylor binary pulsar PSR1913+16. There is a rate of
decrease of the orbital period [$P\sim(3.2\pm0.6)\times10^{-12}ss^{-1}$] as
well as precession of the order of $4^{0}$ per year. Observations are in
excellent agreement with the GR predictions. See, J.H. Taylor, L.A. Fowler and
P.M. McCulloch, "Measurements of general relativistic effects in the binary
pulsar PSR1913+16", Nature (London) \textbf{277}, 437-440 (1979).

$^{7}$To repeat, the "force" is not the force in the mechanical sense.
Nonetheless, a material particle acted on by a mechanical force of the same
form will follow the same null track. See Ref.16 below for the citation of an
interesting example.

$^{8}$Sir A.S. Eddington calculated the bending of light rays round a massive
object by assuming that the space around the sun is filled with a medium with
a refractive index $n=(1-2m/r)^{-1}\approx1+2m/r$. Incidentally, he was also
the first to experimentally observe such a bending in 1919 during the solar
eclipse in Sobral, Brazil. See, A.S. Eddington, \textit{Space, Time and
Gravitation} (Cambridge University, Cambridge, 1920), reissued in the
Cambridge Science Classics Series, 1987, p.109. However, F. de Felice (Ref.10)
reports that Einstein himself was the first to suggest the idea of equivalent
refractive medium.

$^{9}$Among a number of works in this direction, we list only a few, just to
give an idea:\ J. Plebanski, "Electromagnetic waves in gravitational fields",
Phys. Rev. \textbf{118}, 1396-1408 (1960). B. Bertotti, "The luminosity of
distant galaxies", Proc. R. Soc. \textbf{294}, 195-207 (1966). F. Winterberg,
"Deflection of gravitational waves by stellar scintillation in space", Nuovo
Cim. B \textbf{53}, 264-279 (1968). See also Ref.10 for a collection of other references.

$^{10}$F. de Felice shows that the equivalence of a gravity field with the
refractive medium can be successfully employed as a method of investigation:
F. de Felice, "On the gravitational field acting as an optical medium", Gen.
Rel. Grav. \textbf{2}, 347-357 (1971).

$^{11}$Because of the general coordinate covariance of Einstein's GR, its
solutions can be freely expressed in any coordinate system we like. Such a
freedom, however, does \ not affect the unique observable predictions of GR.
This important point is further illuminated in some recent papers. See, Ya.B.
Zel'dovich and L.P. Grishchuk, "The general theory of relativity is correct",
Sov. Phys. Usp. \textbf{31}, 666-671 (1988), and references therein. T. Ohta
and T. Kimura, "Coordinate independence of physical observables in classical
tests of general relativity", Nuovo Cim. B \textbf{106}, 291-305 (1991). A.N.
Petrov, "New harmonic coordinates for the Schwarzschild geometry and the field
approach", Astron. Astrophys. Trans. \textbf{1}, 195-205 (1992). We would
particularly recommend these papers to the advanced graduate students and
researchers in GR.

$^{12}$T. Sj\"{o}din, in \textit{Physical Interpretations of Relativity
Theory}, edited by M.C. Duffy (British Soc. Philos. Sc., London, 1990), pp
515-521. Sj\"{o}din also obtains the light orbit equations starting from
Rindler's prescription. See, W. Rindler, Essential Relativity (Springer, New
York, 1977), especially secs. 8.3 and 8.4.

$^{13}$The reason for fixing the $\theta=\pi/2$ place is that $\partial
/\partial\varphi$ is a Killing field on the entire manifold while
$\partial/\partial\theta$ is not.

$^{14}$Eq.(16) is the classical optics equation. See, A. Mar\'{e}chal,
\textit{Optique G\'{e}ometriqu\'{e} G\'{e}neral\'{e}}, edited by S.
Fl\"{u}gge, Handbuch der Physik, Vol.34 (Springer, Berlin, 1956), p.44. M.
Born and E. Wolf, \textit{Principles of Optics}, 2nd ed. (Pergamon Press, New
York, 1964), pp. 121-124.

$^{15}$A number of light orbits has been plotted in S. Chandrasekhar,
\textit{The Mathematical Theory of Black Holes} (Oxford University, Oxford,
1983). See also the classic treatise: L.D. Landau and E.M. Lifshitz,
\textit{The Classical Theory of Fields} (Pergamon, New York, 1975), pp. 313-316.

$^{16}$W.M. Stuckey, "The Schwarzschild black hole as a gravitational mirror",
Am. J. Phys. \textbf{61}, 448-456 (1993). The possibility of a boomerang
shaped light orbit is interesting in its own right. We can say that the path
of a real massive boomerang around a central mass is also obtained by the same
force law in mechanics. The total energy, however, need not be zero.

$^{17}$The detailed calculations appear in C. M\O ller, The Theory of
Relativity, 2nd ed. (Oxford University, Oxford, 1972), pp.498-501. The
variable speed of light is given by $c^{\ast}=c_{0}(1+2\chi/c_{0}^{2})^{1/2}$,
where $\chi=-GM/r^{\prime}$. Purely spatial curvature is represented by the
metric $ds^{2}=c_{0}^{2}dt^{2}-\left(  1-\frac{2m}{r^{\prime}}\right)
^{-1}dr^{\prime2}-r^{\prime2}(d\theta^{2}+\sin^{2}\theta d\varphi^{2}).$
M\O ller expresses the null trajectory in terms of $c^{\ast}$ and the metric
tensor ($g_{\alpha\beta}$) to demonstrate his result.

$^{18}$Notice that at the Schwarzschild black hole radius $r^{\prime}=2m$,
$n(u^{\prime})=\infty$. Also from Eq.(9), at $r=m/2$, $n(u)=\infty\Rightarrow
c(r)=0$. This result reflects the fact that, to a distant observer, even light
comes to a standstill! To a local observer, however, light always travels at a
speed $c_{0}$.

$^{19}$I.I. Shapiro \textit{et al}, "Fourth test of general relativity:\ New
radar result", Phys. Rev. Lett. \textbf{26}, 1132-1135 (1971). The GR
predicted value of time delay for the Earth-Mercury system is $\sim240\mu\sec$
while the observed value is ($245\pm12$)$\mu\sec$. A remarkable agreement indeed.

\end{document}